\documentclass{iopjournal}
\usepackage{orcidlink}
\usepackage{graphicx} 
\usepackage{subfigure}
\usepackage{gensymb}
\usepackage{amsmath}
\usepackage{xcolor}
\usepackage{hyperref}
\usepackage{lineno}
\usepackage{ulem}
\usepackage{gensymb}
\usepackage{lineno}
\begin{document}
% \linenumbers
\articletype{Paper} %	 e.g. Paper, Letter, Topical Review...

\title{Non-contact temperature sensing and optical
scatter in cryogenically-cooled crystalline silicon}

\author{Disha Pankaj Kapasi$^{1,*}$\,\orcidlink{0000-0001-8189-4920}, Travis Counihan$^1$, Kevin Kimes$^1$, Alina Sendowsky$^1$, Alexandra Gruson$^1$, Daniel Martinez$^1$, Michael Rezac$^1$, Kevin Silverio$^1$, Sean Sorto$^2$, Joshua R. Smith$^1$\,\orcidlink{0000-0003-0638-9670}}

\affil{$^1$The Nicholas and Lee Begovich Center for Gravitational-Wave Physics and Astronomy, California State University, Fullerton, CA, United States of America}

\affil{$^2$Golden West College, 15744 Goldenwest St, Huntington Beach, CA 92647}

\email{dkapasi@fullerton.edu}


\begin{abstract}
Cryogenic crystalline silicon is a promising optical material for next generation gravitational-wave detectors due to its low mechanical loss (a driver of thermal noise), high thermal conductivity and zero coefficient of thermal expansion at 123\,K. In this paper, we report on the design and results of an experiment to measure the temperature of a flat-flat uncoated crystalline silicon sample using a non-contact sensing method relying on interference (optical fringes) made from reflections between the front and back surfaces. We also propose a scheme for implementation of this technique for cryogenic test masses in future gravitational-wave detectors such as Einstein Telescope and LIGO Voyager. Additionally, we characterize the optical scattering of the sample in vacuum at cryogenic temperatures by using a high-sensitivity infrared camera. We measured relatively constant scatter with an average bidirectional reflectance distribution function (BRDF) of $(4.43\pm0.2)\times 10^{-3}\,\mathrm{sr^{-1}}$ at a scattering angle of $\theta_{s} = 7.5\degree$ over the (cold-plate) temperature range of 123\,-\,294\,K and saw no evidence for growth of cryogenic molecular layers (ice) on the sample in our current setup.

% \textcolor{red}{Cryogenic silicon is a promising optical material for next generation gravitational-wave detectors due to its low thermal noise, high thermal conductivity and zero coefficient of thermal expansion at 123\,K. In this paper, we report on the design and results of an experiment to characterize the optical scattering and transmissivity of an uncoated sample of crystalline silicon in vacuum at cryogenic temperatures. The primary results are quantified and imaged changes to the optical scatter versus temperature. We see only small changes (quantify) over the temperature ranges we observed (quantify) and for our vacuum pressure (quantify) and purity, meaning that at these conditions, we do not see signs of ice formation or sample property changes such as damage. We discuss implications for cryogenic gravitational-wave detector optics. A secondary result is the use of optical transmission fringes, made from reflections between the front and back surface of our flat-flat sample, as a non-contact temperature sensor. We discuss how this temperature sensing technique compares to other non-contact methods.}
\end{abstract}

\section{Introduction}\label{sec:intro}
Current room temperature gravitational-wave (GW) observatories (LIGO and Virgo) operate with a primary laser wavelength of 1064\,nm and with mirrors (test masses) made of fused silica with multilayer highly reflective coatings that have low mechanical and optical losses \cite{Granata_2020}.
Thermal noise of the mirror coatings and the test mass suspensions limits the astrophysical range of these observatories~\cite{PhysRevD.111.062002,VirgoCollaboration:25}.
Cryogenic operation of the mirrors and their suspensions offers an additional path---beyond lowering mechanical loss---to reduce thermal noise.
% Overcoming the fundamental limits of coating thermal noise to increase the astrophysical reach of these observatories requires low mechanical loss materials and cooling of optics. 
Previous measurements of fused silica have shown unsuitably high mechanical loss at cryogenic temperatures~\cite{schroeter2007}. Significant research is underway into potential suspension, substrate, and coating materials suitable for 
% long baseline 
cryogenic gravitational-wave detectors.

The Einstein Telescope, LIGO Voyager and Neutron Star Extreme-Matter Observatory (NEMO) are design concepts of future ground-based gravitational-wave observatories that operate with cryogenically cooled crystalline silicon (cSi) test masses to reduce the impact of thermal noise \cite{et2010,Adhikari_2020, nemo}. Thermodynamic fluctuations in the material (substrate and coating) can alter its thickness through thermal expansion and also change the refractive index, both of which are temperature dependent quantities~\cite{middelmann2015, frey}. The KAGRA collaboration currently operates a cryogenic gravitational-wave detector with sapphire test-mass optics~\cite{Akutsu2021}. Recently, cooling of these optics has shown changes in the optical properties of the mirrors due to the formation of cryogenic molecular layers (largely water ice) \cite{PhysRevResearch.1.013008, PhysRevD.99.022003}. Growth of ultra-thin, nanometer-scale ice layers on the optical coating not only changes the refractive index~\cite{Magnozzi_2023} but also leads to excess thermal noise due to internal stress~\cite{PhysRevResearch.1.013008}. Additionally, increased surface roughness as a result of ice growth leads to higher scattered power and lower circulating power in the arm cavities.
% thereby diminishing 
All of these effects can degrade the sensitivity of interferometric GW observatories. 

cSi is a viable alternative option for a test mass material for next-generation GW detectors since it offers highly desirable optical, mechanical, and thermal properties at low temperatures allowing for high-power operation \cite{middelmann2015, rowan2003,  glassbrenner1964}. In this paper, we present a table-top experiment to continuously monitor and study the optical scatter---for 2\,$\mu$m laser wavelength---from cSi at cryogenic temperatures. We also present results that use optical fringes from our transmissivity measurements to predict the temperature of the sample based on its temperature-dependent index of refraction and thermal expansion~\cite{middelmann2015, frey,temp_sense99}. These results provide further data and method development toward the implementation of cryogenic silicon in GW detectors. In section \ref{sec:setup} we introduce the experimental setup and in section \ref{sec:modeling} we discuss the mathematical modeling and results relevant to our experiment. Section \ref{sec:conclusion} highlights future work in this direction.

\section{Experimental procedure}\label{sec:setup}
\subsection{Setup}
\begin{figure}[b!]
    \centering
    \includegraphics[width=0.59\linewidth]{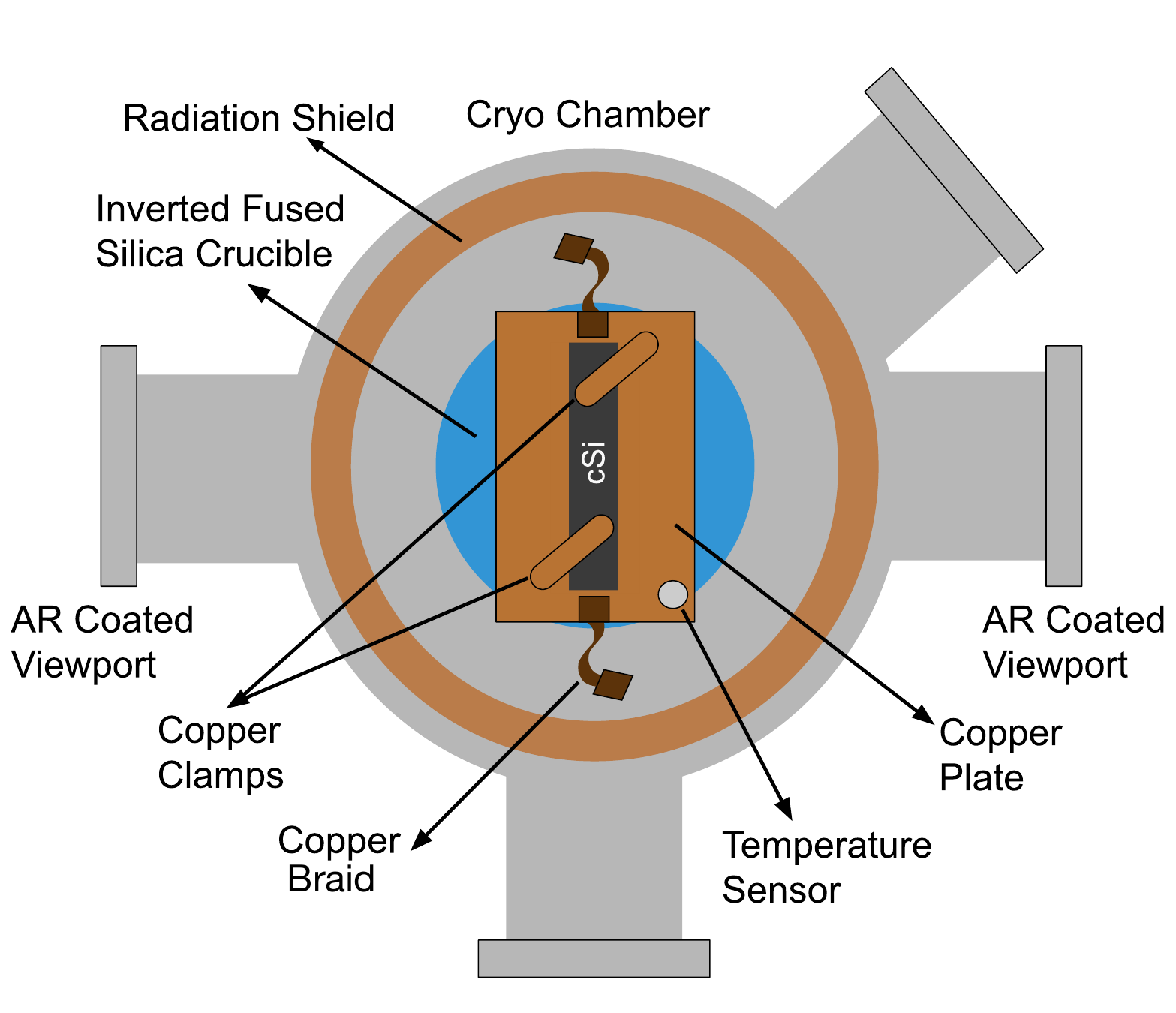}
    \caption{Overhead schematic of the vacuum chamber. A crystalline silicon (cSi) sample rests on top of a copper plate and is held down with pressure from the copper clamps to increase the thermal contact. The copper plate sits on a fused silica crucible (shown in blue). Copper braids link the bottom base plate to the copper plate supporting the sample, thereby increasing the cooldown rate through conduction. With this setup it is possible to cool the copper plate down to 5\,K within 5 hours.}
    \label{fig:setup}
\end{figure}

This section describes our setup to measure the optical scatter and temperature of cSi at cryogenic temperatures. 
We obtained a polished, uncoated, 1-cm-thick crystalline silicon sample (C1-52, cut from disk 8952PT, with a
resistivity of 4360 $\Omega$ cm) from Caltech. The sample was cut from a crystal grown using the magnetic
Czochralski (MCZ) method and originally purchased from Shin-Etsu. Its dimensions are 70 × 30 × 10 mm.
%The dimensions of the sample are 70 x 30 x 10\,mm. 
This flat-flat cSi sample was installed in a vacuum chamber capable of reaching $\sim$5\,K during operation as shown in Fig.~\ref{fig:setup}. The cryo pump from ColdEdge Technologies is a two-stage Gifford McMahon unit where the first stage reaches temperatures of $<$\,40\,K and the second stage reaches 5\,K in $\sim$75 minutes. The second stage is connected to a flexible bayonet that is in thermal contact with the cold plate located inside the vacuum chamber. The interior of the vacuum chamber contains a single layer of copper radiation shield to enhance the rate of radiative cooldown. The optical sample rests on a copper plate that sits on top of a fused silica crucible (fused silica was chosen for its low thermal expansion, so that the sample does not move much during the run, but it has the downside of low thermal conductivity). A small amount of pressure is applied to the sample via two copper clamps to increase the thermal contact between the sample and the copper plate. To improve the sample's cooling efficiency, the copper plate is connected to the cold plate via two copper braids. Silicon diode temperature sensors (DT-670) are installed near the sample (referred to as ``copper plate'') and the cold head (referred to as ``cryostat'') for temperature sensing. The copper plate temperature is controlled by the LakeShore 355 Temperature Controller using a resistive heater as an actuator. The viewports located at 0 and 180 degrees with respect to the sample normal are fitted with antireflection (AR)-coated calcium fluoride windows providing minimal scatter and power loss.

\begin{figure}
    \centering
    \includegraphics[width=0.69\linewidth]{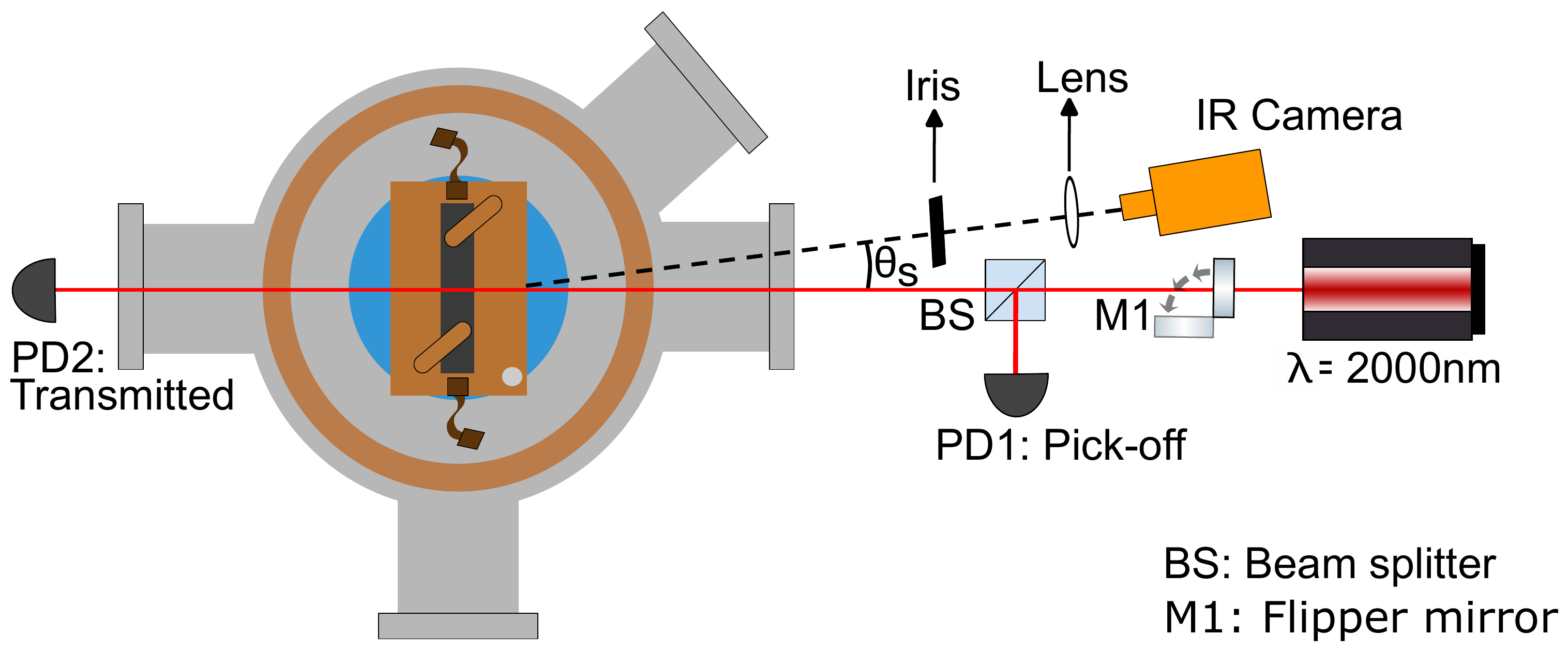}
\caption{Simplified layout of the optical setup. A 1999.9\,nm laser illuminated the sample and an eSWIR Acuros-1280 camera was used to measure the optical scatter from a cSi sample for cryostat temperatures between 5\,K and 295\,K at 7.5$\degree$ scattering angle (with respect to the normal incident beam). The sample was installed in a vacuum chamber (see Fig.~\ref{fig:setup}) and the experiment was controlled using a Virtual Instrument implemented in LabVIEW software.}
    \label{fig:layout}
\end{figure}

Fig.~\ref{fig:layout} shows the optical layout of this experiment. A Thulium-doped fiber laser (Thorlabs LFL2000) was used to illuminate the cSi sample at near normal incidence and 1999.9\,nm wavelength with a maximum optical power of 24.6\,mW at an injection current of 450\,mA. In the path of the incident beam, a beam splitter (Thorlabs BSF10-C) was placed which picks off 2.6\% of the incident power for power monitoring and calibration. Additionally, two integrating sphere photodiode power sensors (Thorlabs S148C) along with calibrated power monitors (Thorlabs PM100D) are used to measure the pick-off (PD1) and transmitted (PD2) optical powers. To image the scatter on the surface of the sample continuously with high sensitivity, an infrared camera (eSWIR Acuros-1280-GigE-001) was installed at an angle of 7.5$\degree$ with respect to the incident beam.

A Virtual Instrument developed in LabVIEW software is used to automate system control and data acquisition for the experiment. A proportional-integral-derivative (PID) loop supervised by the Lakeshore controller uses the copper plate sensor as its error signal and allows for user-designed temperature control profiles with various ramp rates and set points and a precision of 0.01\,K. 

A typical experimental run involves the following procedure. The sample is cooled down at a steady rate of 0.5\,K/min to a setpoint which is held for a specified amount of time, typically several days, to observe possible changes in scatter. During the entire run, incident and transmitted laser power, copper plate and cryostat temperature, elapsed time, and images are recorded to observe any changes in the cSi sample. We take pairs of consecutive ``bright" (laser on) images and ``dark" (laser off) images of the optic with an exposure time of 30\,ms at one-minute intervals for noise subtraction. To avoid switching the laser on and off for the bright and dark images respectively, which can introduce variations in the laser beam properties, a motorized flipper mirror (Thorlabs MFF101) mounted with a neutral density filter (Thorlabs NENIR60A) is used to temporarily block the laser beam from illuminating the sample.  At the end of the run, the chamber is warmed up to room temperature at the same 0.5\,K/min rate. After each run, the data is analyzed using a custom MATLAB script. The images are processed, with dark images subtracted from bright images, and synced with the measured laser power and temperature data to produce videos of the optical scatter and plots of optical scatter and transmissivity versus time. To obtain accurate scatter measurements, a calibration of the infrared camera is necessary (discussed below in Sec.~\ref{sec:calibration}).

\section{Modeling and results}\label{sec:modeling}

\subsection{Optical fringes for temperature sensing}
\label{sec:fringes}

Mounting a temperature sensor on an optic, e.g., in gravitational-wave detectors, can induce mechanical stress, damage the optical surface, and lead to excess vibrational or thermal noise. 
The temperature of the optics in current room-temperature GW detectors can be monitored without contact in several ways. First, temperature monitors in the buildings~\cite{Nguyen_2021}, such as Resistive Temperature Detectors, are located near to the optic locations but outside of the vacuum system. These provide a crude temperature estimate for the optics. Second, measurements of the frequencies of the internal vibrational modes of the mirrors can be compared with Finite Element Analysis models to accurately extract the temperature profile (to the mK) of the optic (and constrain the laser power absorbed by the mirror coatings)~\cite{PhysRevD.103.022003}.
This method relies on using the sensitive GW detector to read out the mirror vibrational modes. Sometimes, LIGO also uses the spring sag of the quadruple pendulum for temperature sensing~\cite{LIGO_T1400749}.    
Third, forward-looking infrared (FLIR) cameras can provide thermal imaging (e.g.,~\cite{Rosauer:25}). However, these cameras are effective only somewhat below room temperature and ineffective at cryogenic temperatures, where the blackbody radiation of the optic is very low (owing to its $T^4$ dependence).
% For example, they work down to sample temperatures of -20C. https://movitherm.com/product/flir-a50-a70-research-development-kits/?utm_source=Google%20Ads&utm_medium=cpc&utm_term=&utm_campaign=Infrared+Cameras+DSA&utm_content=FLIR+A50+A70+Science+Kits&gad_source=1&gad_campaignid=654511386&gbraid=0AAAAAD8MVJufmAYX07O9DFq1oi6Vm6vKo&gclid=Cj0KCQiAsNPKBhCqARIsACm01fR4tH5Qn2D5ZZYrHOHGIWFS4fZC3VJPGke33D7kbb-G7hGYy2L3kxAaAhMFEALw_wcB

\begin{figure}[t!]    \centering\includegraphics[width=1\linewidth]{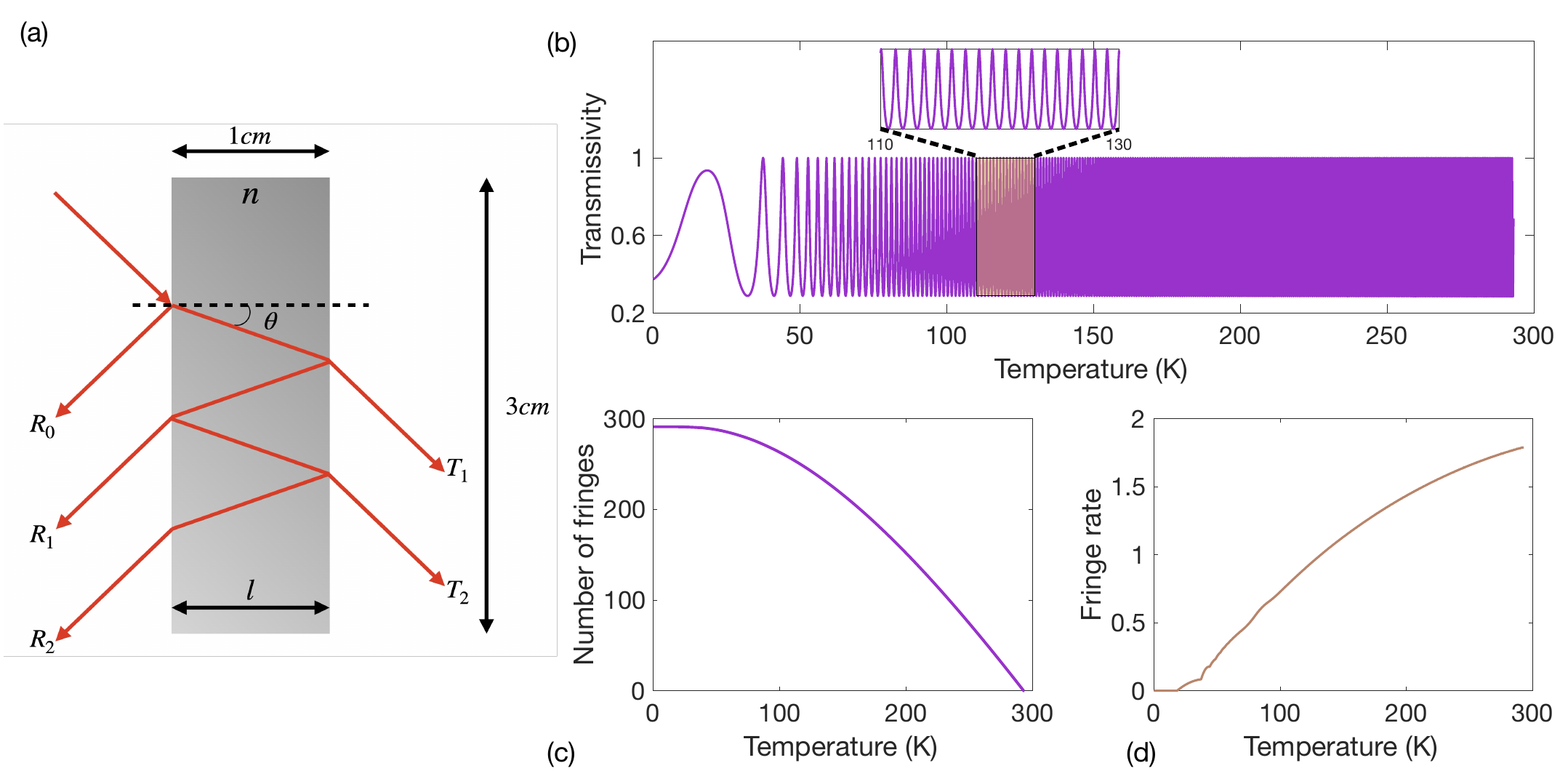}
    \caption{(a) Fabry-P\'{e}rot etalon model of the cSi sample. $R$ and $T$ are reflectivity and transmissivity, respectively, $\theta$ is the angle of incidence, $l$ is the (temperature dependent) sample thickness, and \textit{n} is the (temperature dependent) index of refraction of the sample. (b) Fabry-P\'{e}rot model for the transmissivity of the sample. (c) Cumulative number of fringes starting from zero fringes at room temperature (293.15\,K) and counting upward as the temperature is lowered. (d) Rate of change of fringe count for the cSi sample used in our experiment.}
    \label{fig:fpcav}
\end{figure}

Sensing the temperature of a cryogenic optic without contact is a generally interesting problem. Of the above methods, only the vibrational mode tracking method would work at cryogenic temperatures, and that requires a sensitive length readout. 
LIGO Voyager and NEMO aim to operate at 123\,K where the coefficient of thermal expansion for silicon crosses zero, and the Einstein Telescope (low frequency design, known as ET-LF) is designed to operate $\sim$18\,K. Precisely knowing and controlling the temperature of the cryogenically cooled mirrors in these detectors will be important for maximizing their astrophysical reach. This demands a non-contact, non-invasive cryogenic temperature sensing scheme. 
 
In our experiment, the nearest diode temperature sensor was installed on the copper plate the sample is sitting on (see Fig.~\ref{fig:setup}). 
We expect a temperature difference between the sensor and the cSi sample 
%(specifically, the spot that is illuminated near the center of its largest face) 
due to their separation by several centimeters, the thermal resistance of the interfaces between materials, and the higher radiative heat load on the sample, whose front and back surfaces are exposed to room-temperature blackbody radiation through the viewports and corresponding openings in radiation shield. 
We demonstrate here that a non-contact method based on optical interference can provide a more accurate time-dependent estimate for the temperature of a sample than a nearby diode temperature sensor. Previously, a similar scheme was used to accurately measure the thermo-optic coefficient in silicon down to 5\,K~\cite{komma_thermo-optic}.
% We sought a non-contact method to directly measure the optic’s temperature and compare it with the diode sensor's value. 
% However, any temperature gradient between the sensor and the area of the optic sensed by our laser beam  will lead to discrepancies from the temperature measured by the sensor. 

\subsubsection{Fabry-P\'{e}rot Model}
\label{sec:fpcav}

The cSi optic used in this experiment is a 1\,cm thick polished sample with uncoated parallel surfaces that can be modeled as a Fabry-P\'{e}rot (FP) etalon (or alternatively, as a ``thin" film with vacuum on either side following Hecht equation 9.100~\cite{hecht}).
With the laser incident on the front face of the sample, the reflections between the front and back surfaces create optical fringes as shown in Fig.~\ref{fig:fpcav}(a). The light transmittance of a FP etalon is described by the Airy function written as
\begin{equation}
    T=\frac{1}{1 + F\sin^{2}(\delta/2)}\raisebox{0.5ex}{,}
\end{equation}
 where $F=4 R / (1-R)^2$ is the coefficient of finesse (as opposed to the finesse $\mathcal{F}=\pi\sqrt{R}/(1-R)$) when both surfaces have reflectance $R$ and $\delta$ is the phase difference between the two successive transmitted waves given by
\begin{equation}
    \delta=\left(\frac{2\pi}{\lambda}\right)2nl\cos\theta,
\end{equation}
with $n$ as the index of refraction for silicon, $l$ the sample thickness and $\theta$ the angle of incidence.  

Changes in the temperature of the bulk cSi affect the transmissivity of the etalon primarily via the temperature dependent coefficient of thermal expansion (CTE) and refractive index (RI). 
The changing RI and CTE of silicon leads to phase changes in the multiple-beam interference and thus to intensity modulation of the transmitted light \cite{FPcav1994, temp_sense99}. We note that the number of Airy fringes depends only on $\delta$ and not on $\mathcal{F}$ and thus we have not included losses such as optical absorption in this model.

\begin{figure}[t!]
    \centering\includegraphics[width=0.6\linewidth]{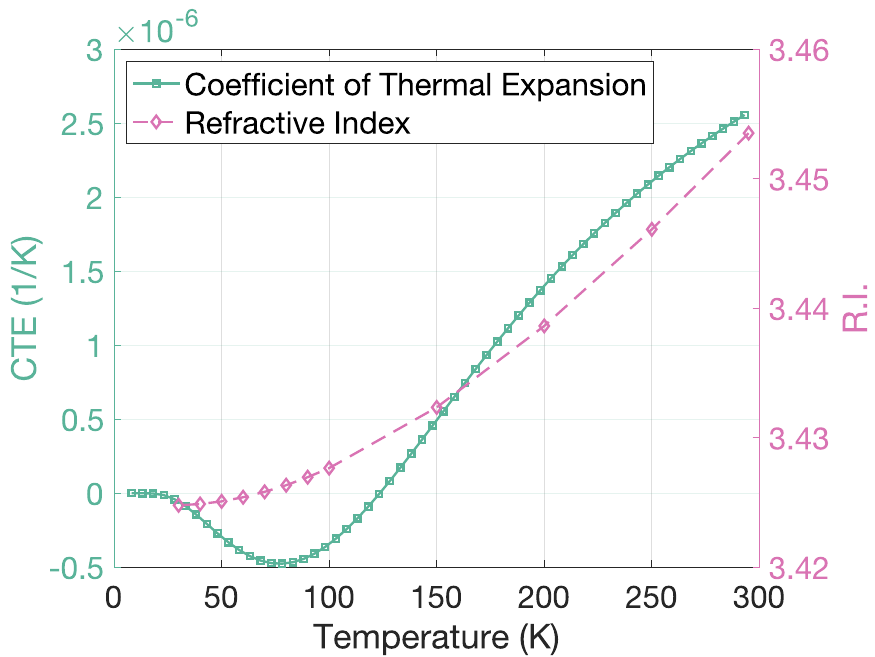}
    \caption{Coefficient of thermal expansion and refractive index of crystalline silicon over the range 0 - 295\,K. Data reproduced from \cite{middelmann2015, frey}.}
    \label{fig:si_prop}
\end{figure}

% Changes in these properties lead to phase changes in the interfering fields of the laser light that is nearly normally incident on the flat-flat faces of the cSI sample  
We make use of the temperature dependent data for CTE from Middelmann~\cite{middelmann2015} and RI from Frey~\cite{frey} as shown in Fig.~\ref{fig:si_prop}. The three samples studied by Middelmann were cut from high purity float-zone-silicon crystals from two separate productions and demonstrated high CTE reproducibility. The prisms used by Frey for RI were reported as ``optical grade" with $>99.999\%$ purity. Our sample was cut from an MCZ-produced cSi crystal and thus, while this is the best available data, some differences are expected. For our model, the values for both the CTE and RI between 0 - 293.15\,K are interpolated at 0.05\,K intervals using the cubic spline method. %In section~\ref{sec:fpcav} we discuss an interferometric non-contact method to calculate the optical fringes from the temperature changes of the cSi sample and \textcolor{blue}{show that the results agree.} %this is a bold statement - check! Using that temperature-dependent RI and CTE data in the FP-model gives the predicted fringes shown in Figure \ref{fig:fpcav}(b).
The fringes in the transmitted light from the FP model for our cSi sample are shown in Fig.~\ref{fig:fpcav}(b). %Notably, the transmissivity here does not take any losses due to absorption into account. 
For counting the optical fringes during a cooling cycle, we can assume zero fringes at room temperature (293.15\,K) and count positively as we reduce the temperature. This is shown in Fig.~\ref{fig:fpcav}(c). From our model, we predict a total of 292 $\pm$ 1 fringes between 0 - 293.15\,K. 
By knowing the number of elapsed fringes we can predict the temperature of our optic. Additionally, the observed passing of a fringe can signal the amount of temperature change corresponding to the fringe spacing at that temperature. Fig.~\ref{fig:fpcav}(d) shows the rate of change of fringes (dN/dT) for the cSi sample used in this experiment.

\subsubsection{Results from fringe counting}\label{sec:fringesresults}

As discussed above, when the temperature of a flat-flat sample such as ours changes, the CTE and RI of the material change, inducing modulation in the transmitted light intensity in the form of optical fringes. To demonstrate this in our setup, we made several runs of cooling our cSi sample to cryogenic temperatures, holding it at a specified temperature (soak), and then warming it back up, all the while, recording the transmitted light power containing the fringes. The details of the various cooldowns and fringe counting results are summarized in Table \ref{table:runs}. During these long runs, several issues caused some data to not be usable (indicated by NA), including power outages, unwanted cryostat warm-up, and equipment errors.

\begin{table}[h!]
\begin{centering}
   \begin{tabular}{|c|c|c|c|c|c|c|}
    \hline
    Run Type & Date & Set point & $\#$ Fringes Ramp Down & $\#$ Fringes Ramp Up\\
    \hline
    Run 1 - CR & 11/15/2024 & 123\,K & 220 & 215\\
    \hline
    Run 2 - CT & 2/26/2025 & 5\,K & 287 & NA\\
    \hline
    Run 3 - CT & 4/25/2025 & 5\,K & NA & 293\\
    \hline
    Run 4 - CR & 4/28/2025 & 123\,K, 18\,K & 226, 289 & NA\\
    \hline
    Run 5 - CR & 7/29/2026 & 123\,K & 226 & NA\\
    \hline
\end{tabular}
\caption{Summary of the cooldowns studied in the CSUF cryogenic experiment (CR: Cold Run; CT: Cold Trap; NA: Data not usable). With reference to zero fringes at room temperature (293.15\,K), the FP model predicts 243$\pm$2 fringes to reach 123\,K and 292$\pm$1 fringes to reach 0\,K.}
\label{table:runs}
\end{centering}
\end{table}

Ahead of the main experimental cooldown (referred to as cold run, CR), a cold trap (CT) run is performed on the apparatus. The purpose of the cold trap is to clean the helium gas and the cryostat and make it free of contaminants. This is achieved by connecting a metal canister housing the cold trap element between the compressor and the manifold and in-line with the helium supply to the cryostat. In this process, the cold trap element (temperature activated charcoal) is cooled to cryogenic temperatures by lowering it into the metal canister containing liquid nitrogen. The temperature activated charcoal captures contaminants that come into contact with it at low temperatures. This process is done for at least 8 hours by continuously running the helium through the cold trap and the cryostat and allowing the vacuum chamber to sit at the minimum temperature ($\sim$5\,K). After the cleaning process is completed, the cold trap is disconnected from the system. The activated charcoal element is then wrapped with heater tape and heated to 100\,C for 24 hours and release the trapped contaminants. 
% \begin{itemize}
%     \item \textit{Cold Trap (CT)} - The purpose of the CT cooldown is to clean the helium gas and the cryostat and make it free of contaminants. This is achieved by connecting a metal canister housing the cold trap element between the compressor and the manifold and in-line with the helium supply to the cryostat. In this process, the cold trap element (temperature activated charcoal) is cooled to cryogenic temperatures by lowering it into the metal canister containing liquid nitrogen. The temperature activated charcoal captures contaminants which comes into contact with it at low temperatures. This process is done for at least 8 hours by continuously running the helium through the cold trap and the cryostat and allowing the vacuum chamber to sit at the minimum temperature ($\sim$5\,K). Once the cleaning process is completed and disconnected from the system, it is connected to a vacuum pump. The activated charcoal is wrapped with heater tape to warm the charcoal (100\,C) for 24 hours and release the trapped contaminants. 
%     \item \textit{Cold Run (CR)} - A CR cooldown is the primary scientific run during which the cSi sample is held at 123\,K over multiple days. At this operating temperature, the CTE for cSi crosses zero leading to vanishing thermo-elastic noise in the material. Since this is highly relevant to the GW community, the objective is to characterize optical scatter over long periods of time. Additionally, the optical scatter is observed around 18\,K which is important for ET-LF.
% \end{itemize}

\begin{figure}[t!]
    \centering
    \includegraphics[width=1\linewidth]{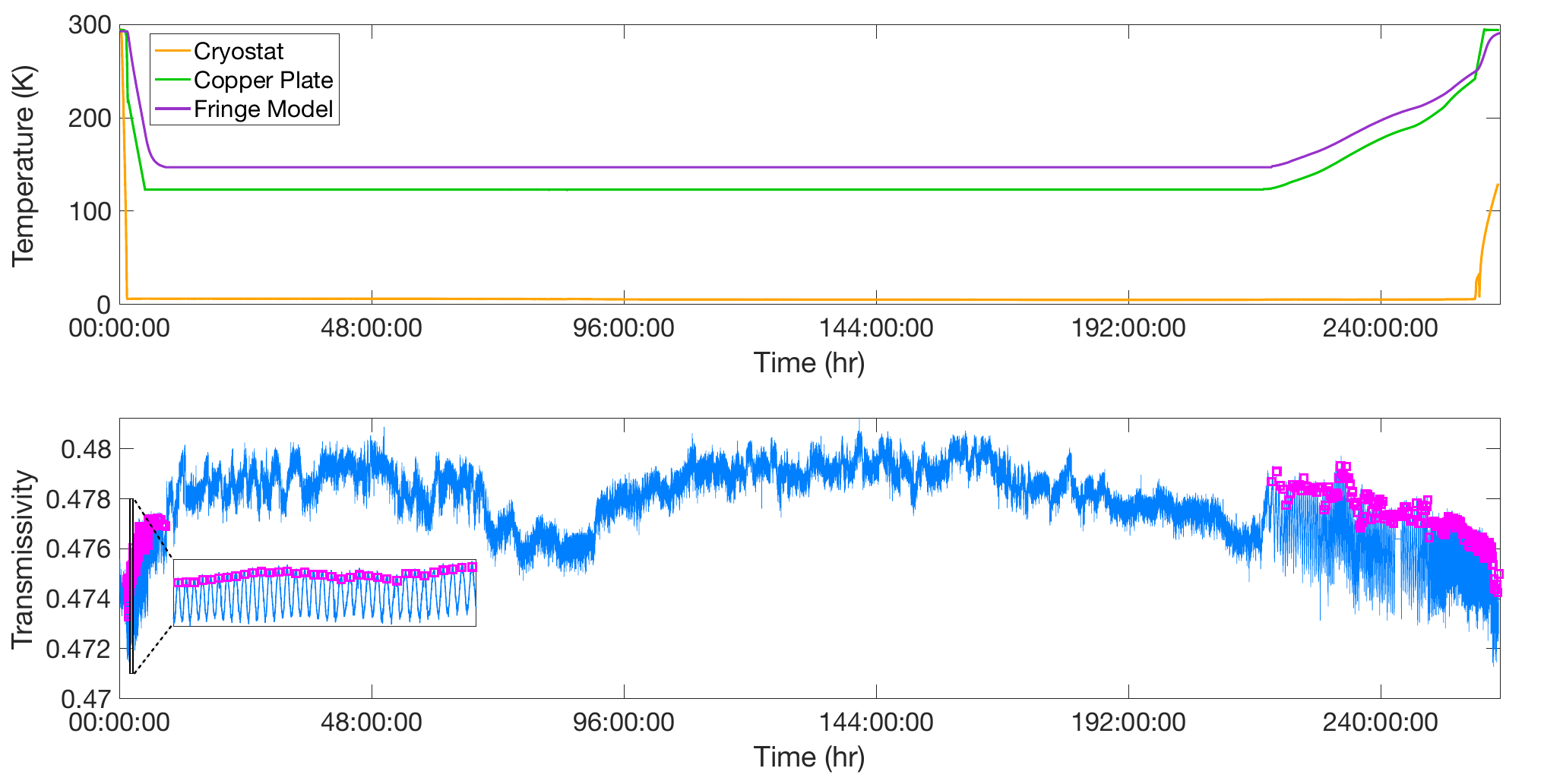}
    \caption{Cooldown from Run 1. The top plot shows the temperature of the cryostat and the copper plate where the temperature sensors are located. The modeled curve based on the fringe counting using the FP model estimates the temperature of the silicon sample to be at $\sim$147\,K while the sensor is at 123\,K. The bottom plot shows periodic changes in the intensity of the transmitted light during the cooldown phase. 220 and 215 fringes were counted during the ramp down and ramp up respectively.}
    \label{fig:fringes_run1}
\end{figure}

%Figure \ref{fig:fringes_run1} and Figure \ref{fig:fringes_run2} are results showing fringe counting from Run 1 (CR) and Run 2 (CT) respectively.

Fig.~\ref{fig:fringes_run1} shows fringe counting analysis from Run 1. The top plot in Fig.~\ref{fig:fringes_run1} highlights the rate of cooldown for the cryostat (yellow) and the copper plate (green) using the temperature sensor data. The temperature of the cSi sample estimated by counting the elapsed fringes (purple) based on the FP model is also shown in this plot. The bottom plot in Fig.~\ref{fig:fringes_run1} shows the varying intensity in the transmitted light due to optical fringes during cooldown and ramp-up. The inset plot shows the optical fringes for a small section during cooldown. The pink squares in this plot highlight the location of the peaks determined using a custom MATLAB script. In Run 1, 220 fringes were counted during cooldown. Although the FP model predicts 243$\pm$2 fringes to reach 123\,K, the lower fringe count on the transmitted photodiode during cooldown suggests that the temperature of the sample is $\sim$147\,K and not 123\,K as recorded by the copper plate sensor. This could be attributed to the lag in the thermal response due to the finite separation between the center of the sample and the sensor on the copper plate, thereby leading to a temperature gradient. The additional heat radiation on the sample due to the direct field of view to radiation at 300\,K from the viewports increases the thermal load on the sample, thereby leading to a higher temperature as indicated by the optical fringes. 

\begin{figure}[t!]
    \centering
    \includegraphics[width=0.99\linewidth]{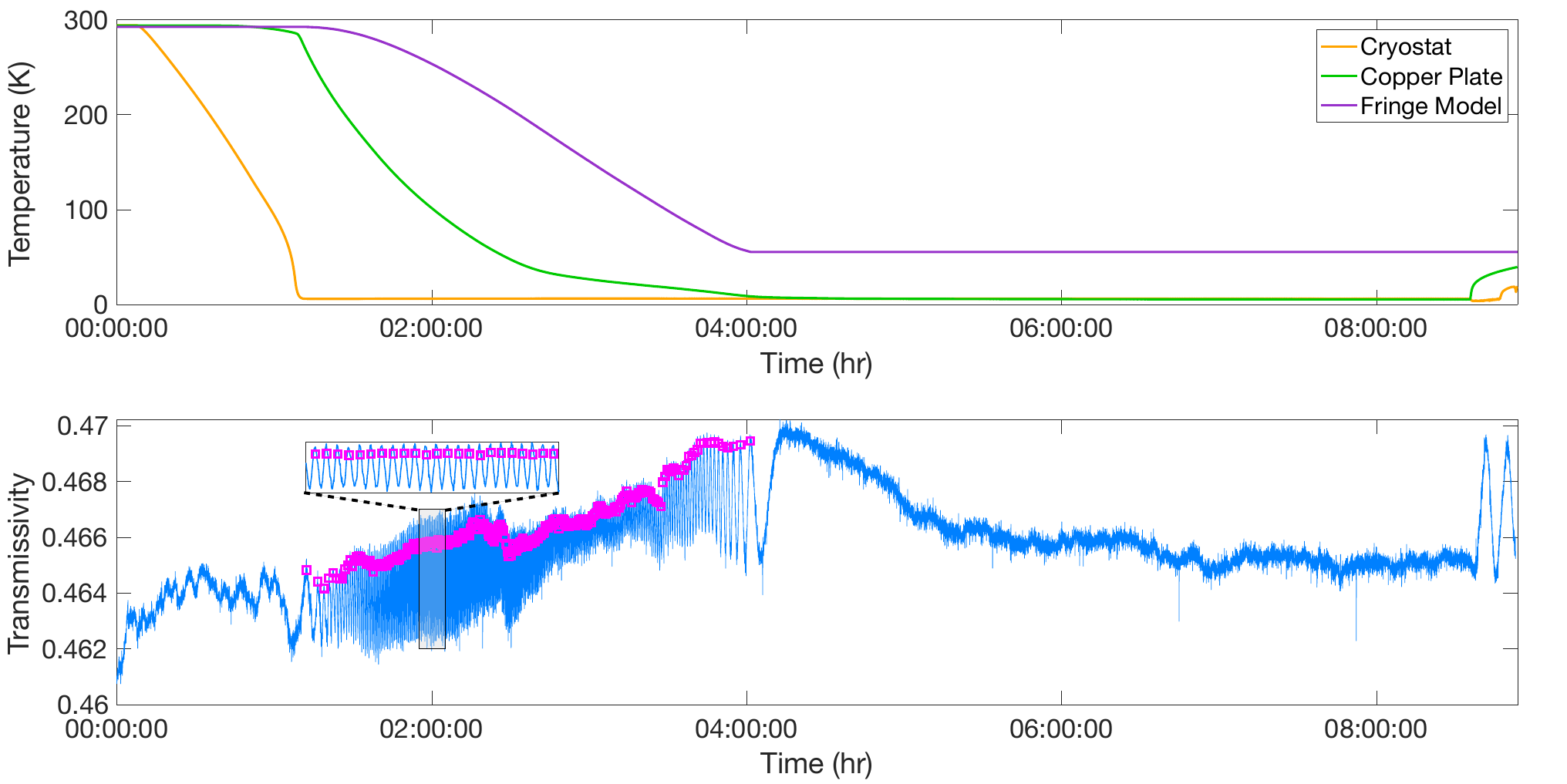}
    \caption{Cooldown from Run 2. The top plot shows the temperature of the cryostat and the copper plate where the temperature sensor is located. The modeled curve based on the fringe counting using the FP model estimates the temperature of the silicon sample to be at $\sim$55\,K while the sensor is at $\sim$5\,K. The bottom plot shows periodic changes in the intensity of the transmitted light resulting in 287 fringes during the cooldown.}
    \label{fig:fringes_run2}
\end{figure}

Fig.~\ref{fig:fringes_run2} shows fringe counting from Run 2. In this cooldown, the cryostat was turned on and for the first 60 minutes, the sample temperature was held at ambient room temperature using a PID control loop. After 60 minutes, the sample was allowed to cool down at a
fixed rate of 0.5\,K/min. 
A total of 287 fringes were counted during cooldown, corresponding to a steady state temperature of $\sim$55\,K. This is quite a large discrepancy from the temperature sensor on the copper plate and indicates that the cooling power is insufficient to overcome the radiation power incident on the sample. 
We also note that below 60\,K the spacing between fringes increases (about 1 fringe every 10\,K) as seen in Fig.~\ref{fig:fpcav}(b)---said differently, the ``resolution" of the fringe counting technique is poorer for temperatures below 60\,K. However, even with that resolution, had the temperature continued to decrease by 10\,K or more toward the set point we would have continued to observe fringes. 

%\textcolor{red}{During the 4/25/2025 cold trap at 02:31:08 there was a ramp up in the chamber temperature from 44.12K to 100.62K resulting in more fringes during the ramp down. Due to the temperature lag between our temperature sensor and optic, we do not know which fringes exactly to exclude from the overall count. The modeled temperature for the ramp down is below zero for this reason. It is uncertain the number of fringes to remove from our count, so an approximation for the temperature in this region is necessary. The fringe count without subtracting fringes is 309.}
% The figures in this section still need to be adjusted to stay in the right location and section: https://www.overleaf.com/learn/latex/Positioning_images_and_tables

%\textcolor{blue}{The transmission is between 0.39 - 0.5 for most data. This can be attributed to - 1. water absorption (see Figure \ref{fig:waterabs}. 2. The upward trend is possibly due to thermal expansion within the chamber leading to pointing error on the transmission PD.}

\begin{figure}[b!]
    \centering
    \includegraphics[width=1\linewidth]{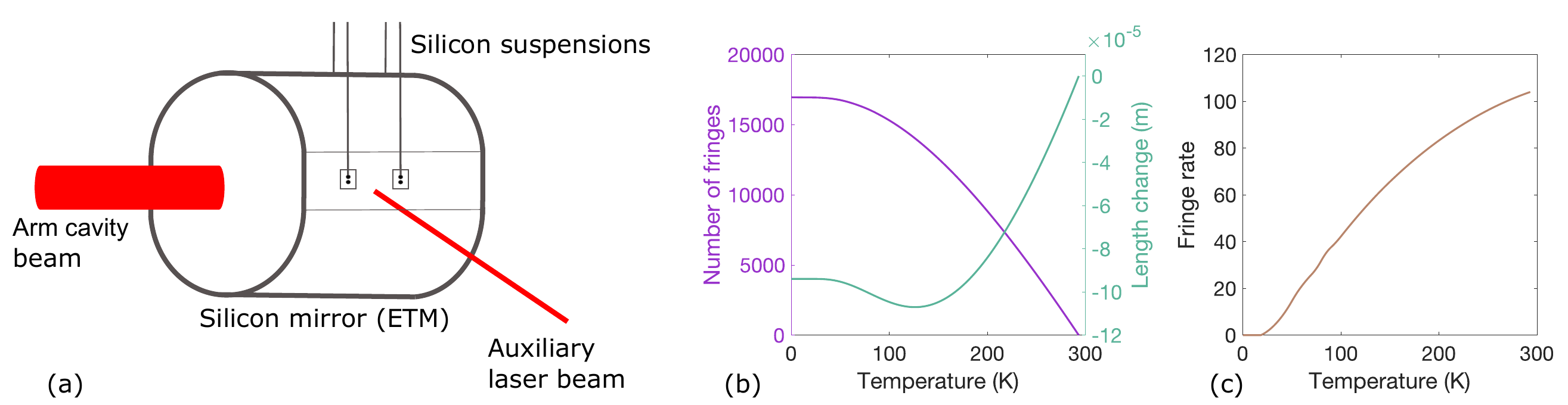}
    \caption{(a) Conceptual design for the implementation of fringe counting for temperature sensing in the ET-LF design. An auxiliary laser at 1550\,nm is incident normally on the barrel flat edge of the mirror. (b) Estimated length change and fringe count for a silicon mirror (test mass) with a 45\,cm diameter. At room temperature the length change is assumed to be zero and the fringes are counted positively with the reduction in temperature. (c) The rate of change of fringes with respect to temperature.}
    \label{fig:etmirror}
\end{figure}

%For the case of Einstein Telescope operating with silicon mirrors at cryogenic temperatures, an auxiliary laser at 1550\,nm can be set up for temperature sensing using fringe counting as shown in Fig.~\ref{fig:etmirror}(a). 
For the case of Einstein Telescope operating with silicon mirrors at cryogenic temperatures, an auxiliary laser at 1550\,nm can be set up for temperature sensing using fringe counting. The test mass optics (mirrors) in gravitational-wave detectors typically have a wedge angle to avoid the etalon effect, so we propose using the flat-flat surfaces that are machined onto the sides of the barrel of the optic as shown in Fig.~\ref{fig:etmirror}(a). %In this proposed scheme, the side edge of the test mass (barrel of the optic) which is flat can be used to probe the mirror at normal incidence. 
A photodiode in transmission (not shown in the figure) is used for monitoring of the transmitted intensity and the fringes can be counted to predict the temperature of the optic. A simple automated tracking algorithm can be integrated for robustly counting the transmission peaks. 
Assuming a diameter of 45\,cm at 292\,K, the estimated length change of the silicon test mass and number of fringes as a function of temperature is shown in Fig.~\ref{fig:etmirror}(b). The much larger dimensions of the Einstein Telescope, combined with the sorter wavelength of the suggested auxiliary laser, are expected to improve the temperature resolution of the fringe-counting technique by $\sim58$ times compared with our experiment (using the material thicknesses of 0.45\,m for ET and 0.01\,m for our experiment and laser wavelengths of 1550\,nm for ET and 1999.9\,nm for our experiment, gives a factor 58.06). This can also be seen by the higher rate of change of fringes in Fig.~\ref{fig:etmirror}(c). 

The proposed fringe-counting method assumes a monotonic change in the optical path length during cooling, in which the direction of phase evolution is known. At steady state, the transmitted intensity alone cannot uniquely distinguish an increasing optical path length from a decreasing one, particularly near the transmission maxima and minima where the first-order sensitivity vanishes. This limitation could be addressed in future implementations by applying a small frequency modulation to the probe laser and synchronously demodulating the transmitted signal. Combining the demodulated signal with the DC transmission would provide additional phase information, allowing the direction of optical-path changes to be determined and improving fringe tracking during non-monotonic temperature variations.

\subsection{Optical Scattering}
\label{sec:scatter}
% In any optical system, scatter is unwanted. 
Optical scattering, caused by coating and surface roughness and imperfections, is often an unwanted source of loss in experiments~\cite{Stover}. 
% From O4 sensitivity paper: "During the break between O4a and O4b, both Livingston ETMs were cleaned using First Contact. After this cleaning, wide-angle scatter from each mirror decreased around 20–40%. Correspondingly, the power-recycling gain at Livingston increased by about 15%."
Optical scatter can decrease the sensitivity of GW detectors by reducing the circulating power in the arm cavities~\cite{PhysRevD.111.062002,Smith_Zucker_2012}, degrading and decohering quantum squeezed states of light~\cite{PhysRevD.90.062006,PhysRevX.13.041021}, and by introducing phase noise through light that is accidentally scattered back into the main beam~\cite{Soni_2021,Flaminio_2010}. Additionally, for cryogenic GW detectors, any ice growth on the surface of the coatings can increase the surface roughness (and thus the scattering) and change the optical properties of the mirrors \cite{PhysRevResearch.1.013008,PhysRevD.99.022003}. Therefore, characterizing the scatter at cryogenic temperatures of potential GW optical materials---such as cSi---is necessary for the implementation of cryogenics in future GW observatories. 

The Bidirectional Reflectance Distribution Function (BRDF) can be used to quantify the scatter from a reflective surface when a light beam is incident on it. Under the assumption that the scatter is from the surface and not the bulk of the material, we can define BRDF as \cite{Stover},
\begin{equation}\label{eq:brdf}
    \begin{centering}
        \text{BRDF} \approx \frac{P_s / \Omega_s}{P_i \cos \theta_s},
    \end{centering}
\end{equation}
where $P_{s}$ is the total scattered power over solid angle $\Omega_{s}$, $P_{i}$ is the total incident power,  $\theta_{i}$ is the angle of incidence, and $\theta_{s}$ is the (polar) scattering angle, measured with respect to the normal of the optic (in the plane of incidence such that the azimuthal scattering angle $\phi_{s}=0$). %Often, BRDF is measured as a function of $\theta_{s}$ (and sometimes also $\phi_{s}$), and presented as angle-resolved scatter. 

\subsubsection{Camera Calibration}
\label{sec:calibration}
The camera calibration process is adapted from Ref.~\cite{ARSpaper} and requires measuring the scattered power from a strongly scattering source using both the low-noise camera and a calibrated power meter at the same scattering angle. For the strong scatterer, we used a sample of the material Spectralon, which is a nearly Lambertian scatterer, meaning that it evenly diffuses the light incident on it in all backward directions and gives a value for % bidirectional reflectance distribution function (BRDF) 
BRDF close to 1/\(\pi\)~\cite{LeeIEEE,Bhandari:11}. 
% By accurately measuring the distances and angles between the sample, camera, and laser, the images recorded on the camera can be calibrated to the known Lambertian power to obtain a calibration factor.
% The Spectralon sample scatters enough to be visible both by our camera and by a less sensitive power meter photodiode installed closer to the imaging optics.
First, we placed a calibrated power meter near enough to the Spectralon sample to measure the scattered power at the scattering angle $\theta_{s} = 7.5\degree$. Using this, the incident power, and the known solid angle (determined by its aperture and distance from the sample) we calculated the sample's BRDF (using Eqn.~\ref{eq:brdf}) at that angle and checked that it was close to the value expected from the literature~\cite{LeeIEEE,Bhandari:11}.  
Then, at the same scattering angle, we used the low-noise camera to take pairs of bright and dark images of the Spectralon. Because the camera is much more sensitive than the power meter, even at a greater distance from the Spectralon it was necessary to install a low-transmissivity neutral-density filter in front of the camera to avoid saturation.
Finally, relating the response at the same angle for the calibrated photodiode and the camera, taking into account the different distances (and thus solid angles) and the camera exposure time, provides a factor that can be used to calibrate the camera values into BRDF. The improvements to the calibration analysis with respect to~\cite{ARSpaper} are listed below.  
\begin{itemize}
    \item Region of interest: The analysis now includes cascading multiple regions of interest (ROIs) which enclose the main scattering area illuminated by the laser beam. 
    The y-intercept obtained from a linear fit of the sum of the camera counts enclosed by each ROI as a function of the ROI area in pixels is then used to calculate BRDF. 
    % Then, the y-intercept determined from the linear fit of the enclosed area of the ROI and sum of camera counts within this area is used to calculate the BRDF from the camera images at that angle. 
    This approach reduces contributions from background noise and other effects that scale with ROI area. 
    % the effect of background noise and other effects that might otherwise increase with ROI area by estimating the scatter enclosed by the innermost ROI.
    \item Elliptical ROI: Using elliptical ROIs instead of rectangles gives a better match to the illuminated scattering spot and removes corners that could bias the BRDF calculations.
    \item Subtracted images: In our calibration script, we
    % create ROIs for the bright and dark images and then 
    subtract the dark image from the bright image to remove common noise and hot pixels. This decreases sources of error for the BRDF. 
    \item Transmissivity of the neutral density filter: We made power measurements of the incident laser beam (directly after the fiber and collimator) before and after one normal-incidence transmission of the neutral density filter (the one used during calibration measurements so as to not saturate the camera) for various incident optical powers. The slope of the linear fit for the transmitted power versus incident optical power provides a more robust power transmission value for the neutral density filter than was used previously.
\end{itemize}

The calibration factor used for reporting the cryogenic scatter (see section \ref{sec:cryobrdf}) at 2000\,nm ($\theta_{s} = 7.5\degree$) is $3.7375\times 10^{-13}$\,W\,s\,counts$^{-1}$\,sr$^{-1}$. %1.5655\times 10^{-13}$\,W\,s\,counts$^{-1}$\,str$^{-1}$.

\begin{figure}[t!]
    \centering
    \includegraphics[width=1\linewidth]{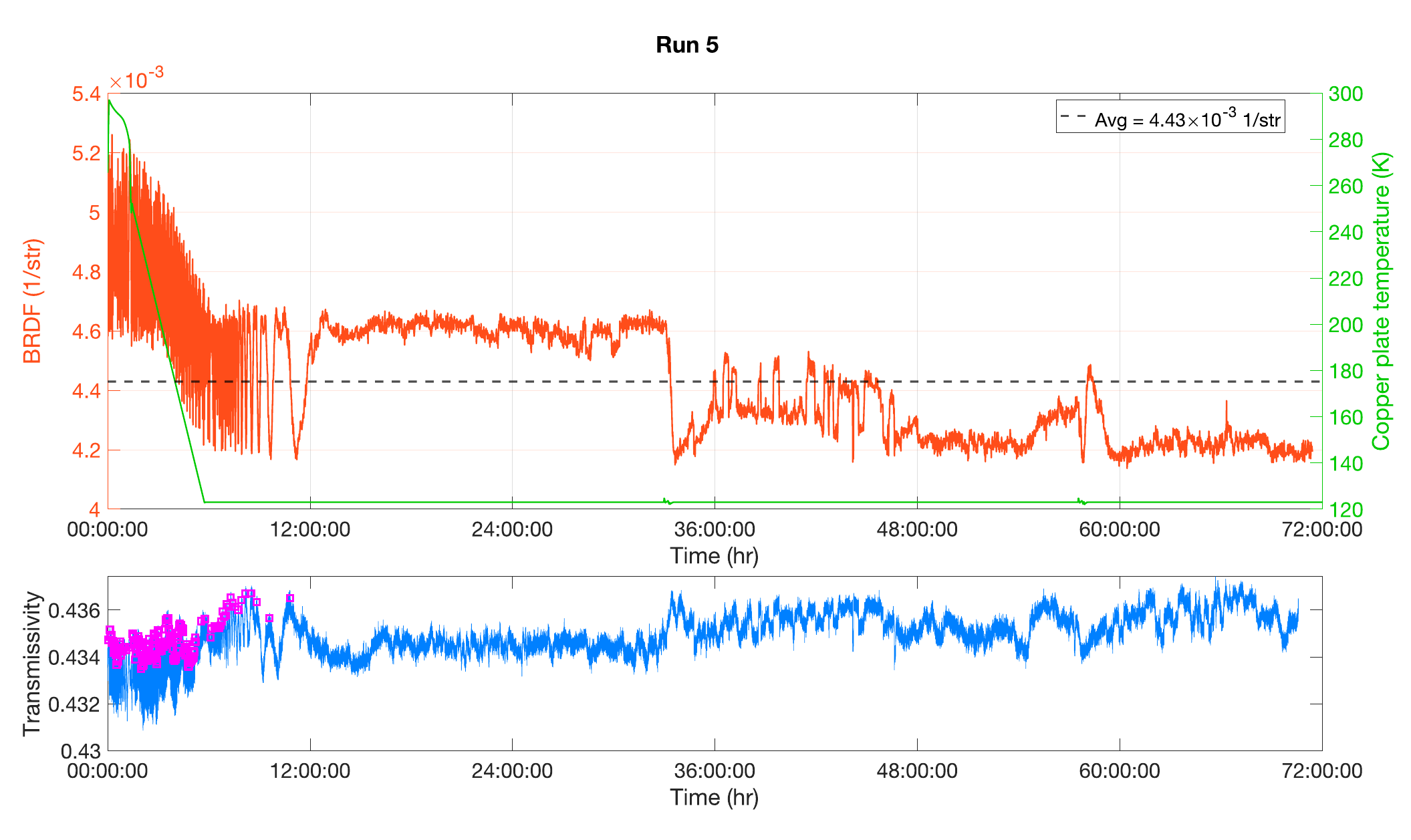}
    \caption{Top plot shows the BRDF from cooldown of Run 5 at a scattering angle (with respect to the sample normal) of $\theta_s=7.5\degree$. The dashed black line indicates the average BRDF. The bottom plot shows the sample's transmissivity which resulted in 226 fringes. The pink squares in this plot highlight the location of the interference peaks.}
    \label{fig:brdfall}
\end{figure}

\subsubsection{BRDF results from cryogenic cSi} 
\label{sec:cryobrdf}
Here we present BRDF results from our cryogenic setup at 2$\mu$m wavelength and with a fixed scattering angle $\theta_{s} = 7.5\degree$ (as in Fig.~\ref{fig:layout}) from Run 5. Ahead of the run, a cold trap was performed and the infrared camera used to capture the bright and dark images was calibrated as described above in sections ~\ref{sec:fringesresults} and ~\ref{sec:calibration} respectively. Additionally, the incident power on the sample is estimated from the pick-off power ($\sim$2.61\%, PD1) which is continuously measured using a calibrated power monitor. 

Fig.~\ref{fig:brdfall} shows the BRDF from Run 5. The dashed black line corresponds to the average BRDF value of $(4.43\pm0.2)\times 10^{-3}\,\mathrm{sr^{-1}}$ where the uncertainty is the standard deviation of the measured BRDF values. In our experiment, we did not see any signs of cryogenic molecular layers (ice) in the BRDF measurements and videos created during the post-processing of the camera images. We attribute this to the fact that the flexible bayonet in the cryostat and the cold plate are significantly colder than the sample and that free gasses and vapor are therefore most likely to adhere to those parts and not to the sample. In a future experiment we plan to force ice to grow on the sample to monitor its growth rate under various environmental conditions and additionally quantify the optical scatter~\cite{Bonilla2023}. The changes observed in BRDF up to hour 12 appear related to coherent effects, i.e., changes in reflectivity of the sample from the etalon effect. The fringes are aliased somewhat in the BRDF data because the corresponding images are only taken once per minute (whereas the transmitted light is measured once per second). The partial fringes around hour 33 and hour 55 appear to be real consequences of small temperature glitches (seen in the copper plate temperature) that also appear in the transmitted light time series. Outside of these interference-induced BRDF changes, there is no appreciable change in scattering during this 72 hour period.

\subsubsection{Angle-resolved scatter of cSi}
\label{sec:ars}
To gather more information about the sample's scattering and surface roughness characteristics, we also measured BRDF as a function of $\theta_{s}$, at 1$\mu$m, the wavelength available in our imaging angle-resolved scatter setup~\cite{ARSpaper,Padilla:14}. 

The calibration of the angle-resolved scatter results followed a similar procedure as that described for the cryogenic scatter in section \ref{sec:calibration} with the modification that the Spectralon sample was measured over a range of angles between 10\degree--\,70\degree{} in steps of 10\degree. The mean of the calibration factor ($1.6146\times10^{-13}$\,W\,s\,counts$^{-1}$\,sr$^{-1}$) with an RMS spread of 4.1\% over all angles was then used to calibrate the Spectralon images and later report the BRDF from cSi. 

\begin{figure}[b!]
    \centering
    \includegraphics[width=1\linewidth]{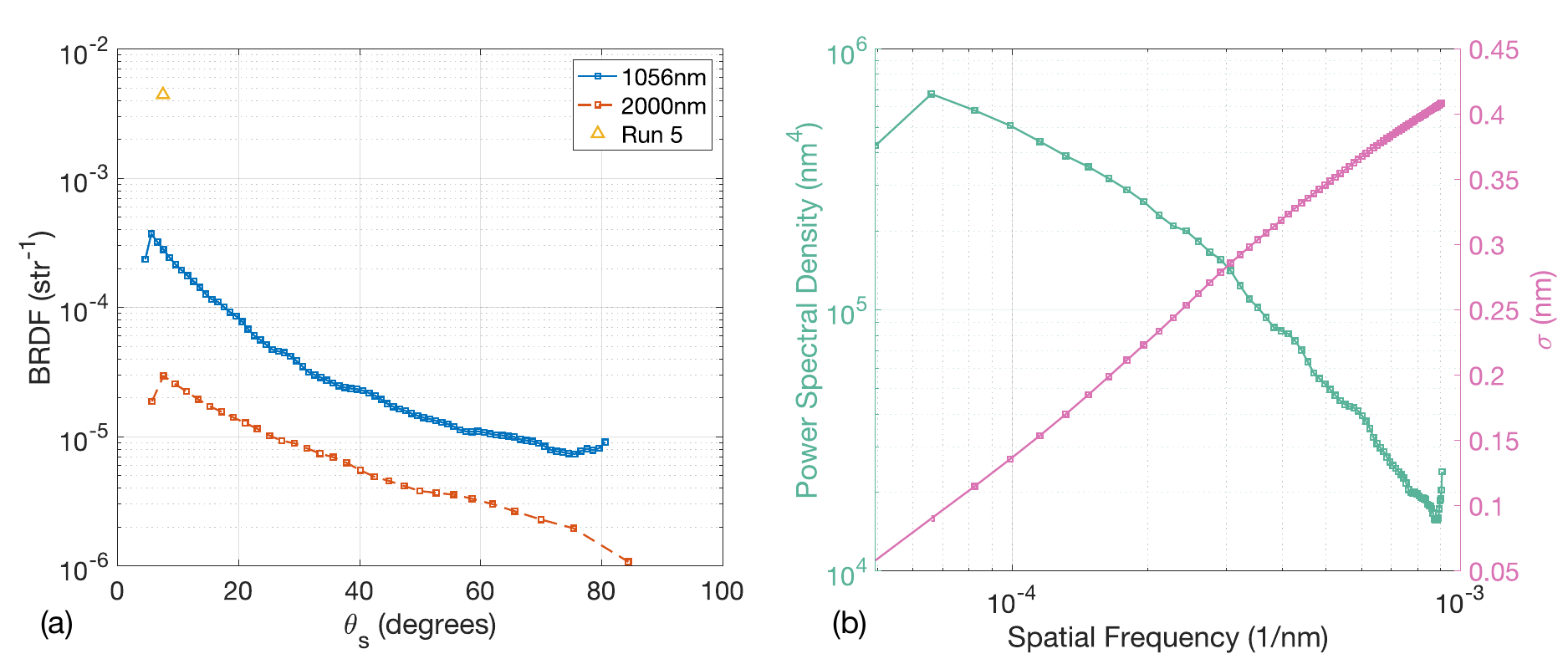}
    \caption{(a) Measured angle-resolved scattering (BRDF) from the polished cSi sample at 1056\,nm (blue line) and predicted BRDF at 2000\,nm (red line, scaled from the 1056\,nm data using Rayleigh-Rice perturbation theory). The triangle indicates the average BRDF value from Run 5 (taken from fig.~\ref{fig:brdfall}). (b) Power spectral density $S(f_n)$ and bandwidth-limited rms surface roughness $\sigma(f_n)$ versus spatial frequency at 1056\,nm.}
    \label{fig:brdf-psd}
\end{figure}

BRDF measurements of the sample's surface scatter 
% measured with our imaging angle-resolved scatterometer~\cite{ARSpaper,Padilla:14} 
at room temperature for scattering angles of 4\degree--\,80\degree{} with 1\degree steps %and with 7.5\degree angle of incidence at room temperature 
are shown in Fig.~\ref{fig:brdf-psd}(a). To scale the measurements to 2000\,nm, we follow the wavelength-scaling procedure described by Stover~\cite{Stover,Trost:13}. For a clean and smoothly-polished surface (with root-mean-squared (rms) roughness much less than the wavelength, $\sigma \ll \lambda$), Rayleigh-Rice perturbation theory relates the measured BRDF to the power-spectral density of the surface roughness~\cite{Stover}, 
\begin{equation}\label{eq:brdfS}
    \begin{centering}
        \mathrm{BRDF} = \frac{16\pi^2}{\lambda^4}\cos{\theta_i}\cos{\theta_s} Q S(f_x,f_y).
    \end{centering}
\end{equation}
Here $f_x=(\sin{\theta_s}\cos{\phi_s}-\sin{\theta_i})/\lambda$ and $f_y=(\sin{\theta_s}\sin{\phi_s})/\lambda$ are the spatial frequencies along the $x-$ and $y-$ directions of the sample. These can be related to the incident and scattering angles  via the grating equation (again with $\phi_s=0$ and here emphasizing the $n$ discrete values), $f_n=\sqrt{f_x^2}=|\sin{\theta_s}-\sin{\theta_i}|/\lambda$.  $Q=\frac{1}{2}\sum_{\alpha}\sum_{\alpha}Q_{\alpha \beta}$ is the polarization factor, which depends in general on the angles of incidence and scatter and the polarization type of the incident $\alpha$ and detected light $\beta$. Our angle-resolved scatterometer utilizes p-polarized light in the plane of the laser beam, giving $Q_{ss}=Q_{ps}=Q_{sp}=0$, so we use for Q~\cite{Germer:97, Trost:13},
\begin{equation}\label{eq:qpp}
    \begin{centering}
        Q_{pp} = \left|
        \frac{(n^2-1)
        \left(\sqrt{n^2-\sin^2{\theta_s}} \sqrt{n^2-\sin^2{\theta_i}} - n^2\sin{\theta_i}\sin{\theta_s}
        \right)}
        {\left(n^2\cos{\theta_i}+\sqrt{n^2-\sin^2{\theta_i}}\right)\left(n^2\cos{\theta_s}+\sqrt{n^2-\sin^2{\theta_s}}\right)}
        \right|^2.
    \end{centering}
\end{equation}
Our cryogenic scatter measurements were made with a laser with random polarization, so we use $Q=(Q_{pp}+Q_{ss})/2$, where~\cite{Germer:97}, 
\begin{equation}\label{eq:qpp2}
    \begin{centering}
        Q_{ss} = \left|
        \frac{(n^2-1)}{\left(\cos{\theta_i}+\sqrt{n^2-\sin^2{\theta_i}}\right)
        \left(\cos{\theta_s}+\sqrt{n^2-sin^2{\theta_s}}\right)}
        \right|^2.
    \end{centering}
\end{equation}
The bandwidth-limited rms roughness of the sample, assuming isotropy, can also be found through appropriate integration of S,
\begin{equation}\label{eq:sigma}
    \begin{centering}
        \sigma = \left[
        \sum_{n=0}^{N-1}2\pi f_n S(f_n) \Delta{f_n}
        \right]^{1/2}.
    \end{centering}
\end{equation}
This sample gives $\sigma=0.41$\,nm at 1056\,nm over the range of scattering angles measured. 
The power spectral density and surface roughness calculated in this way from the BRDF is shown in Fig.~\ref{fig:brdf-psd}(b). 

To complete the scaling, we now plug the surface roughness power spectral density $S$ into Eqn.~\ref{eq:brdfS} and update $Q_{pp}$ for the desired 1999\,nm wavelength, the index of refraction for that wavelength (here at room temperature), and the angle of incidence for the cryogenic experiment (nearly normal), giving the scaled BRDF shown in the left panel of Fig.~\ref{fig:brdf-psd}(a). 

Also shown in Fig.~\ref{fig:brdf-psd}(a) is the average BRDF measured in the cryogenic experiment for Run 5 (discussed in section \ref{sec:cryobrdf}). We expected the cSi BRDF measured in the angle-resolved setup, which uses a 1056\,nm superluminescent diode light source, to be significantly higher than the BRDF measured in the cryogenic experiment, which uses a 1999\,nm laser, due to the inherent scaling of roughness-induced scatter with wavelength (scatter $\propto 1/\lambda^{2}$). However, Fig.~\ref{fig:brdf-psd}(a) shows that the average BRDF from the cryogenic experiment (triangle) is $\sim$150x larger than the predicted angle-resolved scatter at 2000\,nm (red curve) and $\sim$15x larger than the measured angle-resolved scatter at 1056\,nm (blue curve) at the same angle. To understand this discrepancy, we have carefully checked calibrations, camera linearity and scaling and we believe that this discrepancy provides evidence that the assumption that both measurements---predicted angle-resolved scatter at 2000\,nm (red curve) and average BRDF from the cryogenic experiment at 2000\,nm (triangle)---are dominated by surface roughness scattering is incorrect. A possible explanation for this is that the (incoherent source) 1\,$\mu$m angle-resolved scatter measurement is primarily a surface measurement due to strong absorption by silicon at that wavelength while the (coherent source) 2\,$\mu$m cryogenic scatter shows signs of fringing and a more complicated scattering involving at least the front and back surfaces of the sample. 
Another possible explanation is the presence of some specular reflection or ghost beam from the viewport that overlaps with the path of the scattered light that is recorded by the camera (and is therefore challenging to intercept).
Despite its complex behavior and BRDF that exceeds the expectation from surface roughness, the cryogenic scatter measurement at 2\,$\mu$m and its associated images should still serve as a useful diagnostic for the presence of any significant accumulating cryogenic molecular layer---and none was seen.
% , but is likely not a reliable measure of surface-roughness-induced scattering. 

%\textbf{\textcolor{red}{This graph currently shows MORE scatter at 2$\mu$m in the cryogenic experiment than at 1$\mu$m in the ARS experiment, which is not what we would expect if the scatter were all coming from micro-roughness of the same sample. We have not yet found the source of this error, but we are working on it in parallel with the LSC review.}}  \textcolor{red}{As shown above, Figure~\ref{fig:brdf-psd}(a) shows that the average BRDF from the cryo experiment (triangle) is almost 2 orders of magnitude larger than predicted ARS BRDF (red curve) at the same angle. We think this is due to a large calibrtion/analysis error we have not found yet.}

\section{Conclusion}
\label{sec:conclusion}

In this work, we designed and commissioned an optical cryostat operating down to 5\,K and capable of cooling a 70 x 30 x 10\,mm uncoated crystalline silicon sample (MCZ) down to $\sim$55\,K. We demonstrated non-contact temperature sensing for accurately determining the temperature of the sample using interference fringes based on a simple Fabry-P\'{e}rot model. %For cSi sample used in our study, the rate of change of fringes per Kelvin is about 1/K around 123\,K but much lower (about 1/10K) below 60\,K so this technique is more accurate for intermediate cryogenic temperatures.
Accurate implementation of this technique in cryogenic gravitational-wave detectors, such as the Einstein Telescope and LIGO Voyager, will require temperature-dependent refractive-index and thermal-epansion coefficient data for the specific silicon material used by those detectors (e.g., float-zone of MCZ-grown silicon).
% An accurate implementation of this technique for a cryogenic gravitational-wave detector, such as Einstein Telescope or LIGO Voyager, will require temperature-dependent coefficient of thermal expansion and refractive index data for the specific test mass material used (e.g., the specific type cSi---float zone, MCZ, etc). 
The temperature resolution of this fringe counting technique would be much higher in gravitational-wave observatories than in our experiment owing to the much larger dimensions of the test masses with respect to our sample.

We also report optical scatter from our sample at the level of $(4.43\pm0.2)\times 10^{-3}\,\mathrm{sr^{-1}}$ at a scattering angle of $\theta_{s} = 7.5\degree$ over the (cold-plate) temperature range of 123\,-\,294\,K. During this three-day run we saw no evidence for ice growth on the sample. In a future experiment using the same setup, we aim to induce controlled ice growth and study its effect on optical scatter and further develop mitigation strategies for removal of these cryogenic molecular layers using laser-induced desorption.

Together, these results demonstrate that our optical cryostat provides a suitable platform for studying and understanding cryogenic silicon optics (and other materials of interest) under environmental conditions relevant for cryogenic gravitational-wave detectors using non-contact thermometry and in situ scatter monitoring. 

%\textbf{For cSi the rate of fringes per Kelvin is quite high (about 1/K) around 123K but much lower (about 1/10K) below 60K so this technique is more accurate for intermediate cryogenic temperatures.}

%TBA
%
% Each of the commands below will create an unnumbered section with the appropriate heading.
% Remove any sections that are not relevant for your article.
% All sections except suppdata will be removed if the [anonymous] option is used.
% See iopjournal-guidelines.pdf for more information.
%

\ack{We thank our colleagues in the LIGO-Virgo-Kagra Collaboration and in the LIGO Optics Working Group for helpful discussions on the experiment and this paper. We thank Pietro Laudenzi for ideas and input on our fringe counting code and Anamaria Effler and Giles Hammond for ideas on temperature monitoring. We thank Rana Adhikari, Brian Lantz, Christopher Wipf, David Tanner, Yoichi Aso, Takafumi Ushiba, and Sebastian Steinlechner for advice about cryogenic optical measurements.}

\funding{This work was supported by NSF grants 2207998, 2019184, and 2219109 and by Nicholas and Lee Begovich and Dan Black and Family. Separately, Sean Sorto was supported by Project RAISE, U.S. Department of Education HSI-STEM award number P031C210118.}
% This section is a list of funder names and grant numbers

\roles{
\textbf{Disha Pankaj Kapasi:} Conceptualization, equal; Formal analysis, equal; Software, equal; Supervision, equal; Writing - original draft, lead; Writing - review and editing, lead.
\textbf{Travis Counihan:} Data curation, lead; Formal analysis, lead; Methodology, equal; Software, equal; Validation, equal; Writing - original draft, supporting. 
\textbf{Kevin Kimes:} Data curation, supporting; Formal analysis, supporting; Methodology, equal; Software, supporting; Writing - original draft, supporting.
\textbf{Alina Sendowsky:} Investigation, equal; Software, supporting; Writing - original draft, supporting.
\textbf{Alexandra Gruson:} Conceptualization, equal; Investigation, equal; Software, supporting.
\textbf{Daniel Martinez:} Conceptualization, supporting; Investigation, supporting; Software, supporting.
\textbf{Michael Rezac:} Conceptualization, supporting; Investigation, supporting.
\textbf{Kevin Silverio:} Investigation, supporting.
\textbf{Sean Sorto:} Investigation, supporting.
\textbf{Joshua R. Smith:} Conceptualization, equal; Formal analysis, supporting; Funding acquisition, lead; Supervision, equal; Writing - original draft, supporting; Writing - review and editing, supporting.
}
% List author names and the contributions made to the article, using terms from the NISO Contributor Roles Taxonomy (CRediT) https://credit.niso.org

\data{
The data that support the findings of this study are available
within the article.}

\bibliography{references.bib}

\end{document}